\documentclass[twocolumn]{svjour3}          
\usepackage{graphicx}
\usepackage{tikz}
\usepackage{amsmath}
\usepackage{amssymb}
\usepackage{color}
\usepackage{comment}

\definecolor{blueJay}{rgb}{0.541, 0.741, 1}
\definecolor{yellowJay}{rgb}{1, 0.753, 0}
\definecolor{lightBlueJay}{rgb}{0, 0.992, 1}
\definecolor{purpJay}{rgb}{0, 0.125, 0.376}
\definecolor{lightBlue}{rgb}{0, 0.69, 0.941}
\definecolor{midnightBlue}{rgb}{0.1, 0.1, 0.44}
\usepackage[normalem]{ulem}

\journalname{}

\begin{document}

\title{Elastogranularity in Binary Granular Mixtures
}


\author{David J. Schunter, Jr.         \and
            Matthew Boucher		\and
            Douglas P. Holmes$^1$
        }


\institute{David J. Schunter, Jr. \at
              $^1$ Mechanical Engineering, Boston University, Boston, MA, 02215, USA \\
              \email{djsj@bu.edu}      
              }

\date{Received: date / Accepted: date}

\maketitle

\begin{abstract}
Frustration arises for a broad class of physical systems where confinement (geometric) or the presence of a perturbation (kinematic) prevents equilibration to a minimum energy state. By varying the diameter ratio and packing fraction in granular arrays surrounding a slowly elongating elastica, we characterize the resulting {\em elastogranular} interactions taking place in a transitional, amorphous medium. For low number density packings prepared with moderate to large bidispersity, we find the critical injected arclength to elicit jamming follows the same scaling law observed in monodisperse arrays. Beyond the jamming point, the elastica is seen to relax its bending energy within packings with progressively larger diameter ratios towards the shape expected when deforming within more fluid-like media.
\end{abstract}
\section{Introduction}
\label{sec:1}
The packing of granular materials has a well-established history of inquiry, bolstered by theoretical, experimental, and numerical works aimed at understanding these far-from equilibrium systems~\cite{ellenbroek2009,majmudar2007,desmond2009,dagois2012,goodrich2014,coulais2014,morse2017,chaudhuri2010,bitzek2006,morse2014,tighe2014,tordesillas2011}. While individual particles/grains are considered discrete solids, granular amalgams can display a broader range of behavior, transitioning between liquid, glass, and solid-like states~\cite{liu2010,vaagberg2013,mari2009}. This diverse behavior provides direct contrast for situations in which granular matter interacts with a continuum, such as a thin elastic structure. Though in general these coupled interactions are less well understood, they provide a useful connection to many real-world systems. Previous work on root growth has shed light on granular force-chain development and propagation~\cite{kolb2017,kolb2012,whiteley1982,oliva2007,bengough2005}. The study of burrowing bivalves \& crustaceans and of the locomotive strategies in desert dwelling reptiles~\cite{dorgan2015,atkinson2015,maladen2009,young2003} highlights some of the impediments to motion that are specific to moving in and around granular materials. Recent inquiries have aimed to create a general physical framework for these elastogranular phenomena through the analysis of the large elastic deformations of thin rods embedded in both horizontal and vertically oriented granular systems~\cite{kolb2013,mojdehi2016,schunter2018,algarra2018}. With previous investigations either neglecting the role of grain size distribution or limited to monodisperse arrays, questions regarding bidispersity and disruptions to crystalline order, remain open.

\section{Elastogranular Systems}
\label{sec:2}

\begin{figure}
\resizebox{1.0\columnwidth}{!}{\includegraphics{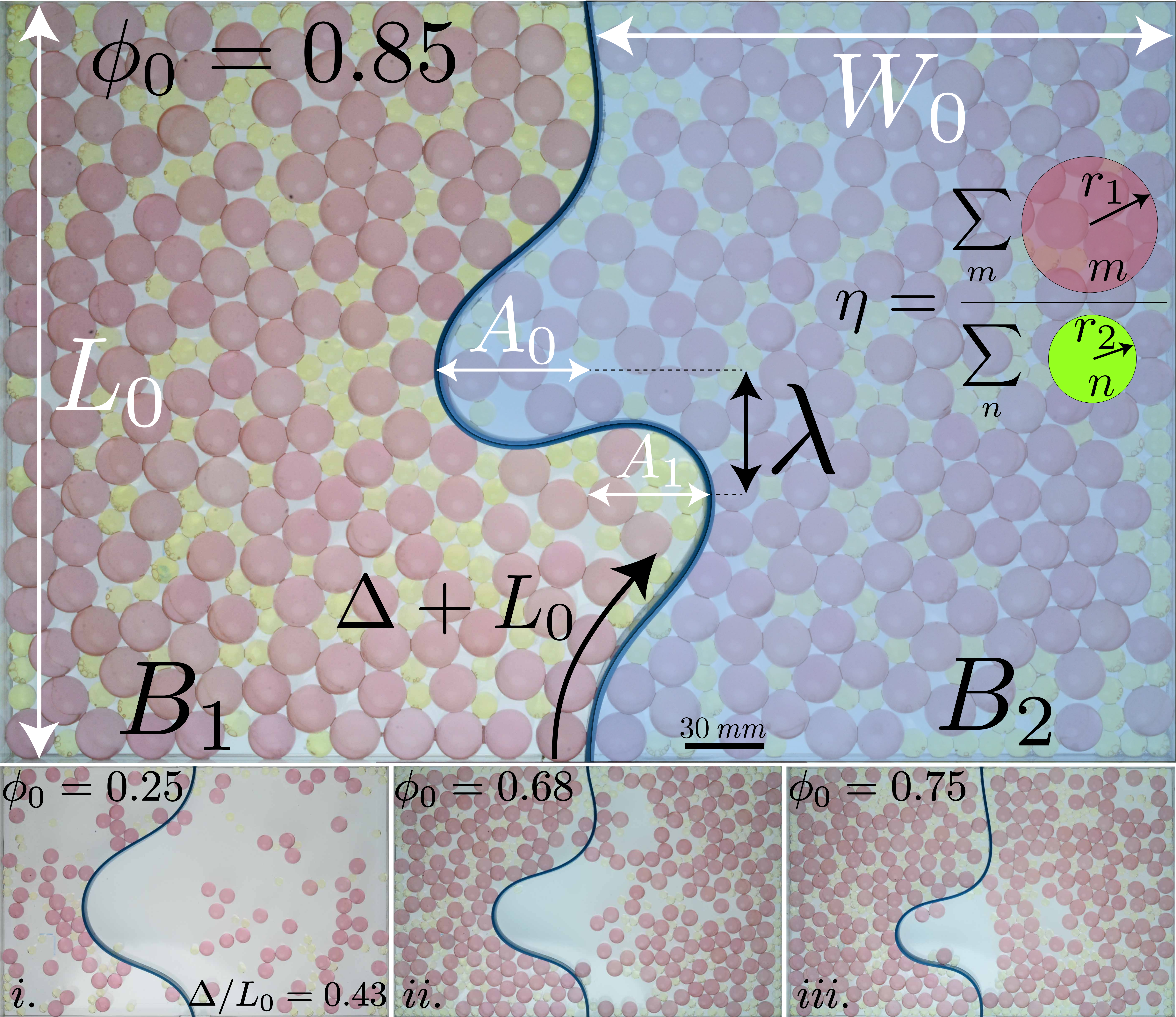}}
\caption{View of experimental set-up. The arclength of a planar elastica is quasi-statically increased by an amount $\Delta$ within granular monolayers at varying diameter ratio $\eta$ (here $\eta=1.9$) and initial prepared packing fraction $\phi_0$. The frames shown above are at the same injected arclength value $\Delta/L_0=0.43$.}
\label{fig1}       
\end{figure}

In this Letter, we consider the buckling and packing of an elastica within a nearly frictionless, bidisperse granular bed. Experimentally, an unbent elastica of initial arclength $L_0$ (equal to the distance separating the clamped/roller boundary conditions that ensure strictly planar deformations and permit additional arclength to enter the system), is confined to deform within 2D arrays of soft hydrogel grains ($\text{M}^2$ Polymer \& MagicWaterBeads). Binary (50:50) mixtures of large $(r_1)$ and small $(r_2)$ radii grains, with diameter ratio $\eta=r_1/r_2$, are randomly placed at equal initial packing fractions $\phi_0$ within the areas $\{B_1,B_2\}$ on both sides of the slender structure [see Fig.~\ref{fig1}]. Here we consider three diameter ratios, with $\eta \in [1.0, 1.2, 1.9]$, prepared over a range of initial packing fractions. At the start of an experiment (for a granular array with a particular $\eta$ and $\phi_0$), we begin increasing the arclength quasi-statically in small increments $\Delta$ $(\sim 0.2\,mm)$, such that the new, current arclength is: $L = L_0 + \Delta$ [Fig.~\ref{fig1} and video S1]. This allows for the observation of both the onset of buckling and characteristic postbuckling morphologies [Figs. 1(i)-1(iii)]~\cite{bigoni2015}.
Using bidispersity as a small perturbation~\cite{kurita2010} to the fragile, hexagonally-packed states that arise in monodisperse arrays ($\eta=1.0$;~\cite{schunter2018}), our aim is to gradually frustrate this global crystalline structure to better understand and characterize elastogranular behaviors in systems where the granular medium acts more like an amorphous solid~\cite{mermin1968}.

\section{The Elastogranular Length Scale}
\label{sec:3}

To better understand the role of bidispersity within elastogranular phenomena, we begin by comparing systems representing two extremes: monodisperse arrays (where $\eta=1.0$) and arrays with moderate to large bisdispersity ($\eta > 1.4$; here $\eta=1.9$)~\cite{kurita2010}. A wide range of experimental packing fractions $\left(0\leq\phi_0\lesssim0.89\right)$ were prepared from which to sample. In general, when the initial packing fraction of an array is below the jamming threshold $\left(\phi_0<\phi_j\right)$, the dominant system effects originate with the deforming elastica. 

In the present experiments, the evolution of the granular contact network, (namely, the reconfigurations taking place due to the lengthening elastica) will not be completely random, as the preparation history of individual packings can allow for small locally crystalline regions to form~\cite{schreck2011,estrada2016,o'hern2003,hamanaka2007,vanhecke2009}. However, given the lack of thermal excitations and the quasi-static nature of arclength injection, we observe no preferential migrations ({\em i.e.} phase separations) of grains towards specific areas of the system~\cite{vaagberg2013,schnautz2005,hamanaka2006} or dominating behavior of one grain size over another within the experimental packings~\cite{kurita2010}.

\begin{figure}
\resizebox{1.0\columnwidth}{!}{\includegraphics{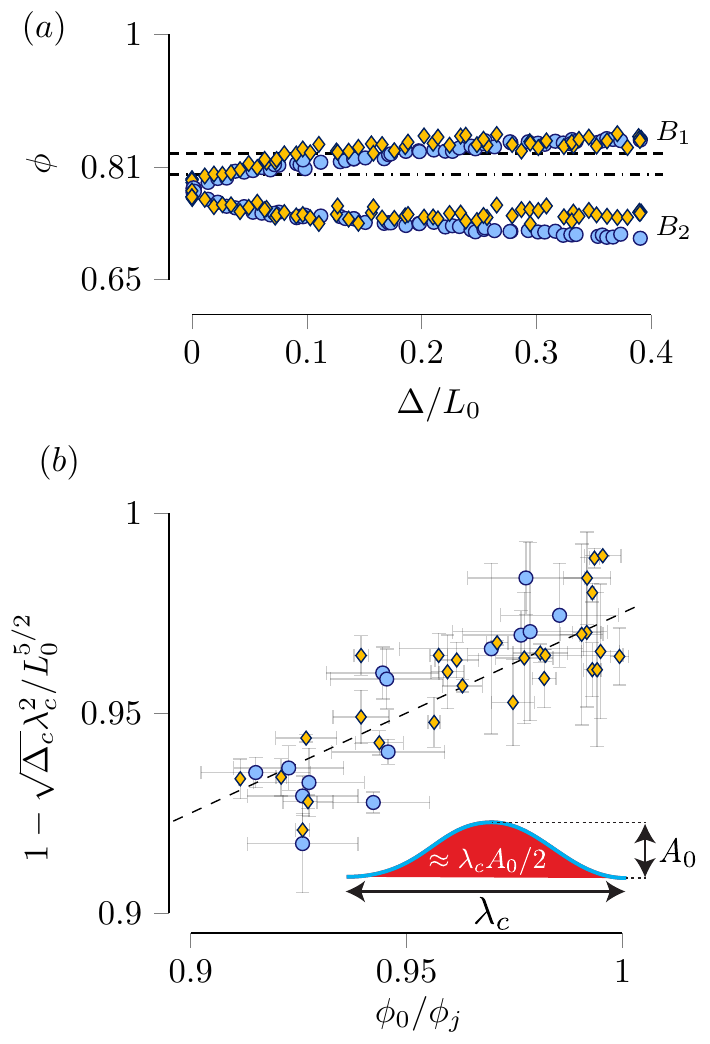}}
\caption{(a) Comparison of pre-jamming behavior for both $\eta=1.9$ (light blue circles) and $\eta=1.0$ (yellow diamonds;~\cite{schunter2018}) arrays. The dashed-dotted (dashed) lines correspond to the critical jamming packing fractions $\phi_j$ for $\eta=1.9$ ($\eta=1.0$) arrays, respectively. (b) The elastogranular length scale $\Delta_c$ is observed to hold in bidisperse arrays ($\eta=1.9$, light blue circles) for $\phi_0<\phi_j$. The dashed line is Eq. $(\textbf{\ref{1}})$, plotted with a slope of 1/2 and the yellow diamonds correspond to experiments with $\eta=1.0$ where the elastogranular length scale was observed, originally discussed in Ref.~\cite{schunter2018}.}
\label{fig2}       
\end{figure}

In the nascent stages of an experiment when $\Delta$ just begins to increase, the thin structure will buckle into a single side of the enclosure (either area $B_1$ or $B_2$),  eventually adopting a mode one postbuckling configuration defined by a primary amplitude $A_0$ and the critical, average half-wavelength $\lambda_c$ measured at low $\Delta/L_0$~\cite{schunter2018}. In pre-jamming arrays $\left(\phi_0<\phi_j\right)$, the elastica will displace grains as additional arclength enters, changing the underlying area available to the granular medium and causing a gradual increase in packing fraction on the side in which $A_0$ grows [Figs. 1(i)-1(iii)]. This side eventually reaches a jammed state at a critical packing fraction $\phi_j$. Critical packing fractions are determined in separate experiments for each value of $\eta$ by placing grains within a rectangular enclosure (as in Fig.~\ref{fig1}) with an adjustable internal area, made possible by a single, rigid actuating wall of length $2W_0$. Taking force measurements with a load-cell (Interface) mounted to the fixed wall opposite the actuating boundary, we determine $\phi_j$ as the point at which the reaction force within the granular array is observed to increase rapidly under continuous quasi-static compression. (See videos S2, S3 and the Supplemental Material in Ref.~\cite{schunter2018} for movies of these compression tests). From these experiments, we find $\phi_j=\{0.8305\pm0.0135,0.8277\pm0.0134,0.7950\pm0.0110\}$ for $\eta=[1.0,1.2,1.9]$, respectively. At comparable initial packing fractions ($\phi_0<\phi_j$), Fig.~\ref{fig2}(a) suggests that even moderate to large ($\eta > 1.4$;~\cite{kurita2010}) bidispersity has little effect on a system's behavior below and on approach to $\phi_j$. Nearly equivalent behaviors are observed between experiments where $\eta=1.9$ (light blue circles) and $\eta=1.0$ (yellow diamonds;~\cite{schunter2018}).

To induce jamming in monodisperse packings, we showed that the critical injection length $\Delta_c$, or the {\em elastogranular} length of this system, can be determined by approximating the area removed from one side of the array as being triangular in shape [inset, Fig.~\ref{fig2}(b)]~\cite{schunter2018}. The primary amplitude is connected to the wavelength by the so--called {\em slaving} condition~\cite{davidovitch2012,paulsen2018}, {\em i.e.} $A_0/\lambda \sim (\Delta/\pi^2L)^{1/2}$, which provides a convenient way to approximate the area consumed by the elastica's deformation as a function of injected arclength. For a fixed number of grains, the critical injection length is found by comparing the area consumed by the elastica with the area that needs to be removed to induce jamming, yielding the relation~\cite{schunter2018}: 
\begin{equation}
\label{1}
\frac{\Delta_c}{L_0}\sim \left(\frac{L_0}{\lambda_c}\right)^4\left(1-\frac{\phi_0}{\phi_j}\right)^2\,.
\end{equation}
We plot this equation in Fig.~\ref{fig2}(b) using values of $\Delta_c$ experimentally measured in arrays where $\eta=1.9$ (light blue circles), along with data from~\cite{schunter2018} (yellow diamonds) for arrays where $\eta=1.0$. Below jamming, it seems that individual packings of bidisperse grains follow the same scaling law as for monodisperse grains. These results, along with the evolution of $\phi$ as a function of $\Delta/L_0$ in Fig.~\ref{fig2}(a), indicate that below jamming the elastica is sensitive to the initial packing fraction~\cite{schunter2018}, but insensitive to variability in the grain size ratio.

\section{Elastogranular Frustration}
\label{sec:4}

In contrast to the situation below jamming, the behavior of packings prepared above $\phi_j$ tends to be dictated by the highly dense granular array~\cite{schunter2018}. In the monodisperse case, the elastica is seen to localize deformations within a diamond-shaped ``lozange" region, the boundary contour of which is set by the tendency of the grains towards hexagonal packing, and the length of which is governed by a characteristic granular length scale $\lambda_c$, reflecting the extent to which forces originating with the thin structure may diffuse out into the medium~\cite{schunter2018}. In these configurations the elastica is kinematically frustrated. With the introduction of bidispersity, where $\eta>1.0$, we begin to observe a qualitative change in the way curvature $\kappa(s)$ localizes along the curvilinear coordinate $s$ of the arclength of the lengthening elastica.

\begin{figure*}
    \centering
    \begin{minipage}[b]{0.5\textwidth}
        \centering
        \includegraphics[width=\textwidth]{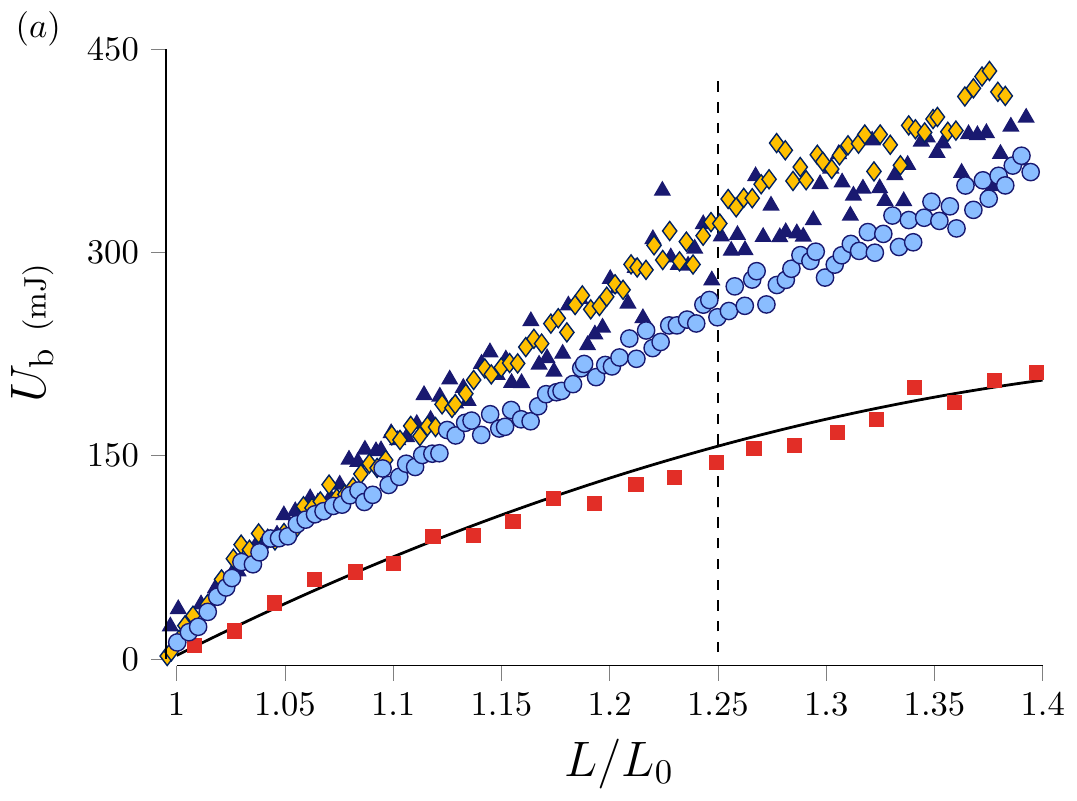}
    \end{minipage}
    \begin{minipage}[b]{0.45\linewidth}
        \centering
        \includegraphics[width=6cm,height=3.25cm]{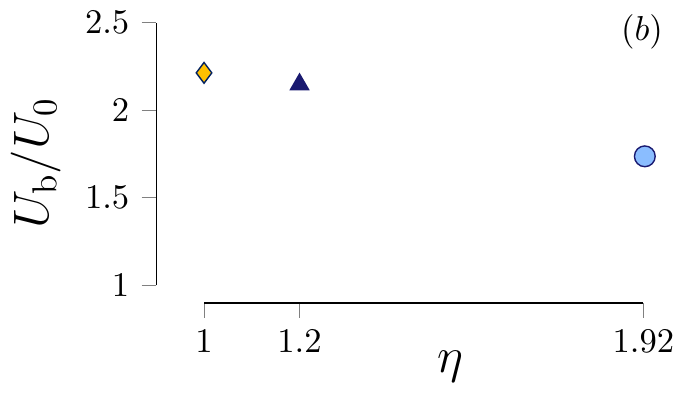}        
        \vspace{\baselineskip}
        \vspace{.15cm}
         \includegraphics[width=5.5cm,height=2.25cm]{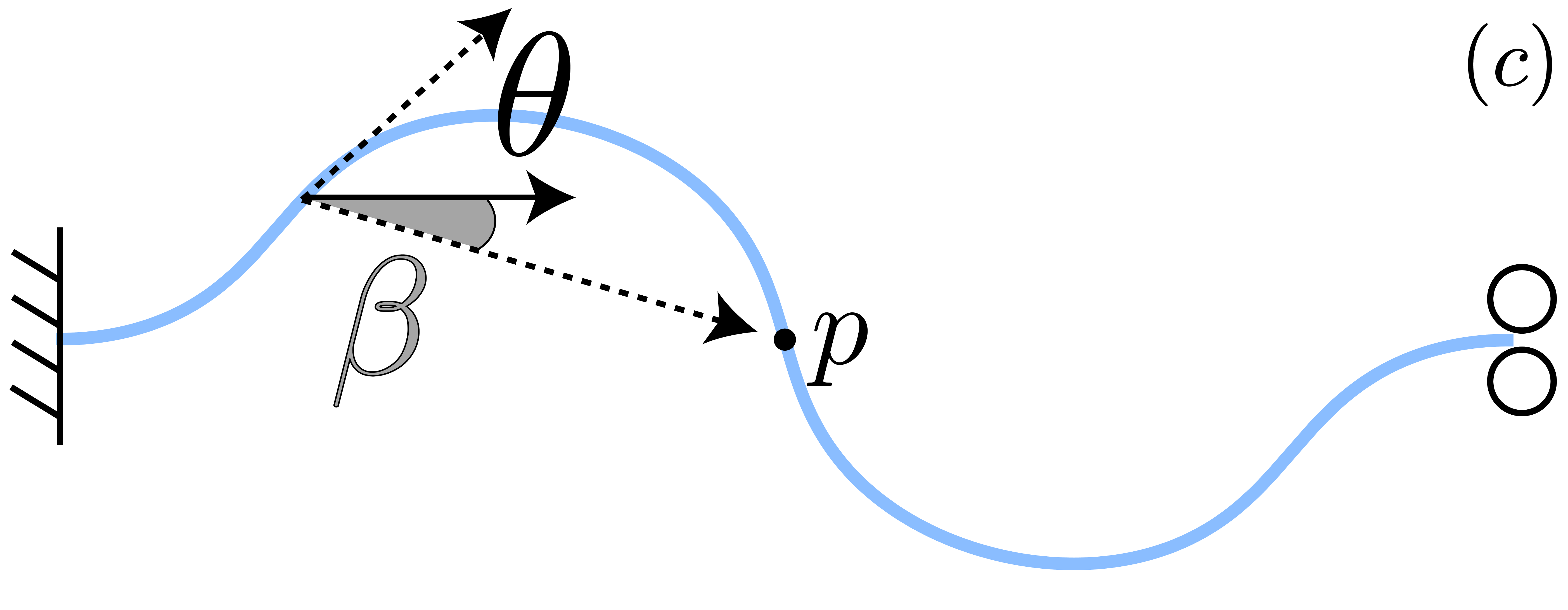}
    \end{minipage}
\caption{(a) Comparison of bending energy values. The evolution of the bending energy $U_{\text{b}}$ as a function of injected arclength is shown for the free antisymmetric elastica (red squares), $\eta=1.0$ (yellow diamonds), $\eta=1.2$ (dark blue triangles), and $\eta=1.9$ (light blue circles), where $\phi_0>\phi_j$. The solid line is the numerical solution for the ideal case of a doubly-clamped, antisymmetric elastica when rescaled by the appropriate beam material and geometric properties. (b) For a fixed injection length $L/L_0=1.25$ [dashed vertical line, (a)] in arrays with equal $\phi_0$, the dimensionless bending energy, normalized by the free-case bending energy $U_0$ [red squares, (a)], is observed to decrease as the diameter ratio $\eta$ becomes larger. (c) Defining the inclination angles $\{\theta,\beta\}$, inflection point $(p)$, and clamped-roller boundary conditions, along with the numerically calculated elastica profile (at fixed injection length $L/L_0=1.25$).}
\label{fig3}       
\end{figure*}

We quantify this gradual change from over confinement by looking at the elastica's bending energy~\cite{landau1986}: $U_{\text{b}}=\frac{\text{B}}{2}\int_{0}^{L}\kappa(s)^2 \ \text{d}s$, the energy required to bend a structure characterized by a bending rigidity $B=EI$, where $E$ is the material's elastic modulus, and $I$ is the second moment of area (given by $I=h^3b/12$ for a beam of thickness $h$ and width $b$ as measured out of the plane in Fig.~\ref{fig1}). Above the jamming threshold, the dense granular arrays act as an ``effective" elastic medium, confining the elastica with approximately equal pressure contributions from side $B_1$ and $B_2$. Due to this assumed average force balance within the granular bed, we expect the buckling geometry that minimizes bending energy in the beam to be equivalent to the bending energy of an antisymmetric, doubly-clamped elastica. The governing differential equation for this problem is given by:
\begin{equation}
\label{elastica}
\psi''(s)+\gamma^2\sin \psi(s) = 0 \ \ \ \ \forall \ s \in [0, L/2],
\end{equation}
where $\gamma^2=(P^2+R^2)^{1/2}/B$, where $P$ and $R$ are the axial and transverse reaction forces, respectively, at the clamped ends~\cite{bigoni2015}. The angle $\psi(s)=\theta(s) + \beta$ defines the tangent at $s$ relative to $\beta$, which is the inclination of $P$ and $R$ with respect to the axial direction [Fig.~\ref{fig3}(c)]~\cite{bigoni2015}. Equation~\ref{elastica} is valid from $0\leq s \leq L/2$, as its symmetry about the inflection point $(p)$ at $L/2$ reduces the problem to two equivalent clamped--pinned beams. 

The boundary conditions for this reduced problem become $\psi(0)=\beta$ and $\psi(L/2)=0$, and an additional (global) kinematic constraint: $\int_0^{L/2}  \sin(\psi(s)-\beta) ~\text{d}s$, ensures the vanishing of transverse displacements at the clamped end and the inflection point. Solutions to this system of equations are highly non--trivial (see~\cite{bigoni2015} for a clear and detailed explanation), and result in parametric equations for the in--plane displacements, $x(s)$ and $y(s)$. For an elastica that is elongating between two fixed ends, we can multiply the parametric equations by a scalar $\Gamma$ that represents an increment in ``growth'' of the curve~\cite{schunter2018}, such that
\begin{subequations}
\begin{align}
\label{x}
x_g(s) &= \Gamma x(s)\,,\\
\label{y}
y_g(s) &= \Gamma y(s)\,.
\end{align}
\end{subequations}

Unlike the symmetric elastica deformation described in~\cite{schunter2018}, the antisymmetric case depends on two related angles $\psi$ and $\beta$. We used Newton's method for numerical root finding in the commercial software \textsc{Mathematica} to determine $\beta = f(\psi)$. The injected length $\Delta$ is found by numerically integrating the parametric equations~\ref{x} and~\ref{y} for a range of $\Gamma$-values. Finally, the bending energy of these curves is found by numerically integrating the square of the arc curvature [solid line, Fig.~\ref{fig3}(a)]. By measuring the experimentally observed bending energy in the elastica for representative runs at each value of $\eta$ investigated, we can utilize these numerical results to determine the extent to which variations in $\eta$ may drive the elastica towards this assumed minimal energy configuration [Fig.~\ref{fig3}].

To determine the bending energy $U_{\text{b}}$ for experimental runs, the elastica's deformation profile is extracted from each frame of an image sequence and subsequently discretized using custom image processing code written in MATLAB. Fitting polygons to these discrete points, we can obtain a measurement of the analytic curvature at each point, quantities which are then summed and squared over the elastica's arclength. Indeed, in Fig. 3(a) we observe a gradual decrease in $U_{\text{b}}$ as $\eta$ is made larger, with $U_{\text{b}}[\eta=1.9]$ (light blue circles) less than $U_{\text{b}}[\eta=1.2]$ (dark blue triangles), which in turn is less than $U_{\text{b}}[\eta=1.0]$ (yellow diamonds).

\begin{figure*}
\resizebox{1.0\textwidth}{!}{\includegraphics[height=7cm]{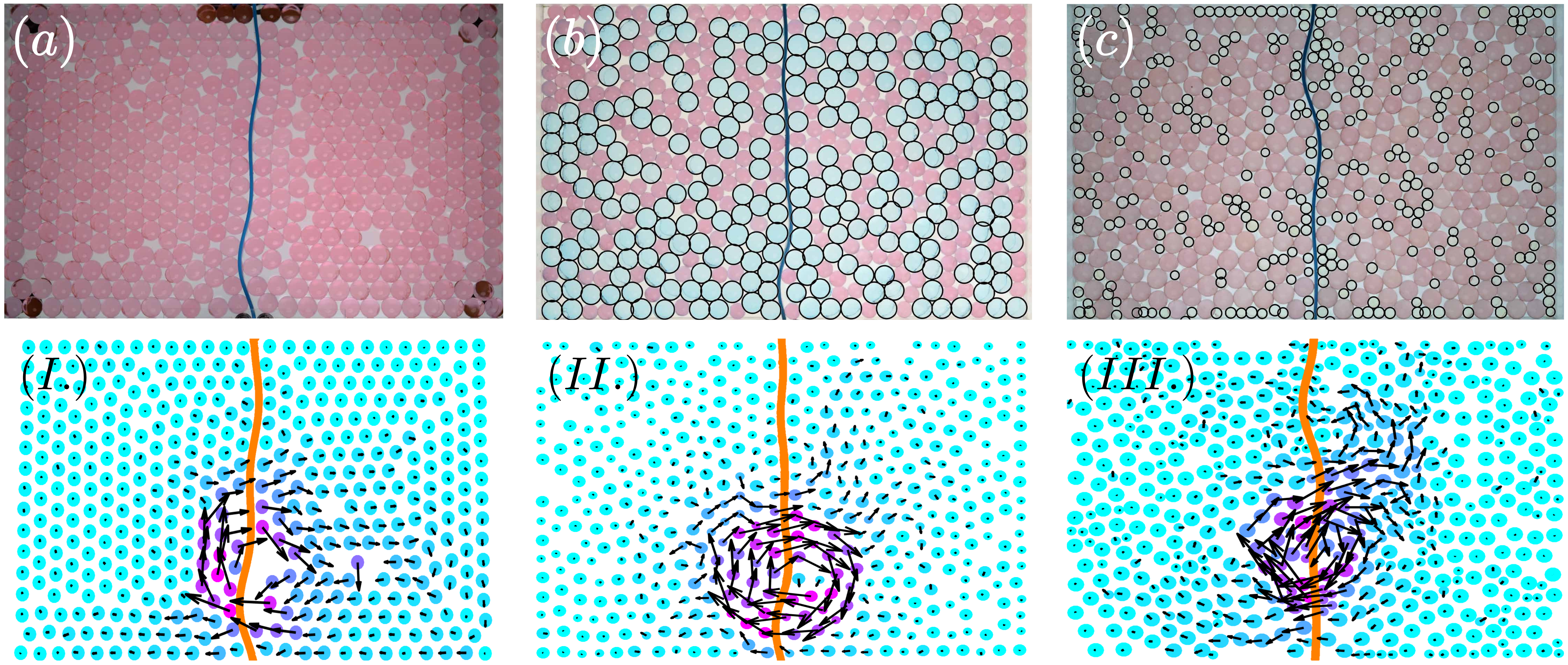}}
\caption{Granular displacement fields at equivalent initial packing fraction ($\phi_0=0.89$) and low injected arclength ($\Delta/L_0\approx0.014$). As the diameter ratio $\eta=1.0$ (a) increases to $\eta=1.2$ (b), $\eta=1.9$ (c), the mobility of grains within the monolayer begins to increase. The crystalline ordering characteristic of $\eta=1.0$ packings (I), which restricts granular motion to small areas of the array, is disrupted by the introduction of bidispersity $[\eta=1.2; (II)]$. At the largest experimentally tested value of $\eta=1.9$, the granular displacement field, no longer confined, is observed along the entire length of the elastica (III). We have outlined one respective size of grain radii used in preparing a specific $\eta$-valued array in (b),(c) to aid visualization.}
\label{fig4}       
\end{figure*}

To experimentally verify the model, we also performed experiments for the idealized case of a free antisymmetric elastica (with bending energy $U_0$) by artificially pinning the midpoint at a given current length $L$, imaging, and analyzing as in the previous experiments [red squares, Fig.~\ref{fig3}(a)]. The numeric and experimental values of the antisymmetric elastica, which serves as our point of comparison by defining an effective continuum limit (where $\eta>>1$) for more ``fluid-like" arrays, are seen to be in excellent agreement. We expect that this continuum limit would also be reached at fixed $\eta$-values if the grain sizes were decreased relative to the elastica thickness. This question could be addressed in a subsequent study.\\
\indent The bidisperse arrays used here lack the global crystalline order found in the $\eta=1.0$ case, allowing for highly localized regions of curvature in the elastica to relax within the granular medium as opposed to being confined within a characteristic region. At comparable injected arclength $\Delta$ and $\phi_0$, along with decreases in bending energy [Fig.~\ref{fig3}(b)], this relaxation manifests as a difference in the mobility of the surrounding granular array. We quantify this by tracking the motion of individual grains within arrays at each $\eta$-value tested [Figs. 4(I)-4(III)]. The elastica must effectively ``fracture" the more solid--like arrays prepared at $\eta \in [1.0, 1.2]$ in order for additional arclength to enter [Figs. 4(I)-4(II)], resulting in a highly localized granular displacement field restricted to a small area of the system. In monodisperse granular arrays, these high-mobility regions have been observed to occur in close proximity to any disruptions of hexagonal ordering~\cite{schunter2018,sausset2008}. This behavior contrasts what we observe at the largest experimentally tested value of $\eta=1.9$: the granular displacement field is no longer confined to a small, characteristic region and grain motion is observed along the entire extent of the elastica's arclength [Fig.~\ref{fig4}(III)]. 
	
\section{Discussion}
\label{sec:5}

It is interesting to note that by introducing geometric frustration (via bidispersity) into the granular medium, we were able to alleviate some of the kinematic frustration present in the confined elastica (observed to adopt a lower energy configuration in Fig.~\ref{fig3}). We speculate the existence of an intermediate range of $\eta$-values and elastica bending rigidity $B$, such that the relative effect each element has on the system balances the other. In this intermediate range, modifications to the rigidity of the thin elastic structure will hypothetically have the same effect as an adjustment to $\eta$. In practice, there are certain physics and engineering scenarios where it may be easier to change the characteristic dimensions of either the elastic structure or the granular medium. Additionally, simple experimental models such as these of coupled, frustrated systems may provide a novel means of investigating equipartition in nonlinear systems~\cite{davidovitch2019}. A thorough understanding of this energy balancing and how elastogranular systems can be ``tuned" will have direct applications to the fields of soil mechanics and civil engineering, and in the development of burrowing robots or steerable needles.

\begin{acknowledgements}
We are grateful for financial support from the National Science Foundation CMMI--CAREER through Mechanics of Materials and Structures (No. 1454153).
\end{acknowledgements}

\section*{Compliance with ethical standards}
\small{\textbf{Conflict of interest} The authors declare that they have no conflict of interest.}

\bibliographystyle{spphys}       
\bibliography{eGPolyDisp}   

\begin{thebibliography}{10}
\providecommand{\url}[1]{{#1}}
\providecommand{\urlprefix}{URL }
\expandafter\ifx\csname urlstyle\endcsname\relax
  \providecommand{\doi}[1]{DOI \discretionary{}{}{}#1}\else
  \providecommand{\doi}{DOI \discretionary{}{}{}\begingroup
  \urlstyle{rm}\Url}\fi

\bibitem{ellenbroek2009}
W.G. Ellenbroek, M.~van Hecke, W.~van Saarloos, Physical Review E
  \textbf{80}(6), 061307 (2009)

\bibitem{majmudar2007}
T.~Majmudar, M.~Sperl, S.~Luding, R.P. Behringer, Physical review letters
  \textbf{98}(5), 058001 (2007)

\bibitem{desmond2009}
K.W. Desmond, E.R. Weeks, Physical Review E \textbf{80}(5), 051305 (2009)

\bibitem{dagois2012}
S.~Dagois-Bohy, B.P. Tighe, J.~Simon, S.~Henkes, M.~Van~Hecke, Physical review
  letters \textbf{109}(9), 095703 (2012)

\bibitem{goodrich2014}
C.P. Goodrich, S.~Dagois-Bohy, B.P. Tighe, M.~van Hecke, A.J. Liu, S.R. Nagel,
  Physical Review E \textbf{90}(2), 022138 (2014)

\bibitem{coulais2014}
C.~Coulais, A.~Seguin, O.~Dauchot, Physical review letters \textbf{113}(19),
  198001 (2014)

\bibitem{morse2017}
P.K. Morse, E.I. Corwin, Physical review letters \textbf{119}(11), 118003
  (2017)

\bibitem{chaudhuri2010}
P.~Chaudhuri, L.~Berthier, S.~Sastry, Physical review letters \textbf{104}(16),
  165701 (2010)

\bibitem{bitzek2006}
E.~Bitzek, P.~Koskinen, F.~G{\"a}hler, M.~Moseler, P.~Gumbsch, Physical review
  letters \textbf{97}(17), 170201 (2006)

\bibitem{morse2014}
P.K. Morse, E.I. Corwin, Physical review letters \textbf{112}(11), 115701
  (2014)

\bibitem{tighe2014}
B.P. Tighe, Granular Matter \textbf{16}(2), 203 (2014)

\bibitem{tordesillas2011}
A.~Tordesillas, G.~Hunt, J.~Shi, Granular Matter \textbf{13}(3), 215 (2011)

\bibitem{liu2010}
A.J. Liu, S.R. Nagel, Annu. Rev. Condens. Matter Phys. \textbf{1}(1), 347
  (2010)

\bibitem{vaagberg2013}
D.~V{\aa}gberg, Jamming and soft-core rheology.
\newblock Ph.D. thesis, Ume{\aa} Universitet (2013)

\bibitem{mari2009}
R.~Mari, F.~Krzakala, J.~Kurchan, Physical review letters \textbf{103}(2),
  025701 (2009)

\bibitem{kolb2017}
E.~Kolb, V.~Legu{\'e}, M.B. Bogeat-Triboulot, Physical biology \textbf{14}(6),
  065004 (2017)

\bibitem{kolb2012}
E.~Kolb, C.~Hartmann, P.~Genet, Plant and Soil \textbf{360}(1-2), 19 (2012)

\bibitem{whiteley1982}
G.~Whiteley, J.~Hewitt, A.~Dexter, Physiologia Plantarum \textbf{54}(3), 333
  (1982)

\bibitem{oliva2007}
M.~Oliva, C.~Dunand, New Phytologist \textbf{176}(1), 37 (2007)

\bibitem{bengough2005}
A.G. Bengough, M.F. Bransby, J.~Hans, S.J. McKenna, T.J. Roberts, T.A.
  Valentine, Journal of Experimental Botany \textbf{57}(2), 437 (2005)

\bibitem{dorgan2015}
K.M. Dorgan, Journal of Experimental Biology \textbf{218}(2), 176 (2015)

\bibitem{atkinson2015}
R.J.A. Atkinson, L.B. Eastman, The natural history of the Crustacea \textbf{2},
  100 (2015)

\bibitem{maladen2009}
R.D. Maladen, Y.~Ding, C.~Li, D.I. Goldman, science \textbf{325}(5938), 314
  (2009)

\bibitem{young2003}
B.A. Young, M.~Morain, Copeia \textbf{2003}(1), 131 (2003)

\bibitem{kolb2013}
E.~Kolb, P.~Cixous, N.~Gaudouen, T.~Darnige, Physical Review E \textbf{87}(3),
  032207 (2013)

\bibitem{mojdehi2016}
A.R. Mojdehi, B.~Tavakol, W.~Royston, D.A. Dillard, D.P. Holmes, Extreme
  Mechanics Letters \textbf{9}, 237 (2016)

\bibitem{schunter2018}
D.J. Schunter~Jr, M.~Brandenbourger, S.~Perriseau, D.P. Holmes, Physical review
  letters \textbf{120}(7), 078002 (2018)

\bibitem{algarra2018}
N.~Algarra, P.G. Karagiannopoulos, A.~Lazarus, D.~Vandembroucq, E.~Kolb,
  Physical Review E \textbf{97}(2), 022901 (2018)

\bibitem{bigoni2015}
D.~Bigoni, \emph{Extremely Deformable Structures}, vol. 562 (Springer, 2015)

\bibitem{kurita2010}
R.~Kurita, E.R. Weeks, Physical Review E \textbf{82}(4), 041402 (2010)

\bibitem{mermin1968}
N.D. Mermin, Physical Review \textbf{176}(1), 250 (1968)

\bibitem{schreck2011}
C.F. Schreck, C.S. O'Hern, L.E. Silbert, Physical Review E \textbf{84}(1),
  011305 (2011)

\bibitem{estrada2016}
N.~Estrada, Physical Review E \textbf{94}(6), 062903 (2016)

\bibitem{o'hern2003}
C.S. O'Hern, L.E. Silbert, A.J. Liu, S.R. Nagel, Physical Review E
  \textbf{68}(1), 011306 (2003)

\bibitem{hamanaka2007}
T.~Hamanaka, A.~Onuki, Physical Review E \textbf{75}(4), 041503 (2007)

\bibitem{vanhecke2009}
M.~van Hecke, Journal of Physics: Condensed Matter \textbf{22}(3), 033101
  (2009)

\bibitem{schnautz2005}
T.~Schnautz, R.~Brito, C.~Kruelle, I.~Rehberg, Physical review letters
  \textbf{95}(2), 028001 (2005)

\bibitem{hamanaka2006}
T.~Hamanaka, A.~Onuki, Physical Review E \textbf{74}(1), 011506 (2006)

\bibitem{davidovitch2012}
B.~Davidovitch, R.D. Schroll, E.~Cerda, Physical Review E \textbf{85}(6),
  066115 (2012).
\newblock \doi{10.1103/PhysRevE.85.066115}

\bibitem{paulsen2018}
J.D. Paulsen, Annual Review of Condensed Matter Physics  (2018)

\bibitem{landau1986}
L.D. Landau, E.~Lifshitz, Course of Theoretical Physics \textbf{7}, 109 (1986)

\bibitem{sausset2008}
F.~Sausset, G.~Tarjus, P.~Viot, Physical review letters \textbf{101}(15),
  155701 (2008)

\bibitem{davidovitch2019}
B.~Davidovitch, Y.~Sun, G.M. Grason, Proceedings of the National Academy of
  Sciences \textbf{116}(5), 1483 (2019)

\end{thebibliography}

\end{document}